\documentclass{elsart}

\usepackage{graphicx}
\usepackage{dcolumn}
\usepackage{bm}

\begin{document}
\begin{frontmatter}
\title{Micropillar resonator in a magnetic field: Zero and Finite temperature cases}
\author{Herbert Vinck-Posada$^{(1)}$, Boris A. Rodriguez$^{(1)}$,
 Augusto Gonzalez$^{(2)}$}
\address{$^{(1)}$Instituto de Fisica, Universidad de Antioquia, AA 1226,
 Medellin, Colombia\\
 $^{(2)}$ Instituto de Cibernetica, Matematica
 y Fisica, Calle E 309, Vedado, Ciudad Habana, Cuba}

\begin{abstract}
In this work, we present a theoretical study of a quantum dot-microcavity system which includes a constant magnetic field in the growth direction of the micropillar. First, we study the zero temperature case by means of a  selfconsistent  procedure with a trial function composed by a coherent photon field and a BCS function for the electron-hole pairs. The dependence of the ground-state energy on the magnetic field and the number of polaritons is found. We show that the magnetic field can be used as a control parameter of the photon number, and we make explicit the  scaling of the total energy with the number of polaritons. Next, we study this problem at finite temperatures and obtain the scaling of the critical temperature with the number of polaritons.   
\end{abstract}
\end{frontmatter}
\maketitle

\section{Introduction}In the last years, the study of many microsystems that confine light and provide the interaction with an active medium has made possible the observation of interesting phenomena such as, for example, the control of the spontaneous emission (Purcell effect)\cite{X1,X2,X3}, or the enhancement of ground state occupation for exciton-polaritons at low temperatures \cite{Dang,Yamamoto2006} (in the search for BEC of polaritons). Many interesting applications to quantum computation \cite{XX1} and other areas are envisaged. 

An exciton-polariton is a quasi-particle composed by an exciton and a photon in the strong coupling regime. These quasi-particles have less reduced effective mass
than atoms or excitons and, consequently, a relatively high condensation temperature is expected. 

In this work, we study the effects of an homogeneous magnetic field, applied in the growth direction of a quantum dot-micropillar system on the energy, the mean number of photons in the microcavity, and the critical temperature for BE condensation of polaritons. Our calculations assume a conserved number of polaritons. Due to the fact that polaritons indeed decay, such an assumption means that their lifetime is much longer than the time required to achieve thermal equilibrium at a fixed number of polaritons \cite{Snocke}.

The paper has been written as follows: In section II, we describe the system and the theoretical model to be used below, in section III, we describe the selfconsistent method used to obtain the ground-state energy. In the next section, we explain the extension of the method to finite temperatures and, finally, in the last section we present numerical results for the ground state energy, mean number of photons and the critical temperature.
 
\section{Theoretical Model}
The heterostructure that we are interested in is a  circular pillar grown by periodic deposition of layers of GaAs and AlGaAs (Bragg mirrors). In the center of this pillar  a $\lambda$-cavity-defect of GaAs is placed. It contains in the middle a set of quantum dots of GaInAs. When  the radius of this micropillar resonator is about  $0.5\mu$m, one can assume that the microcavity operates with a single electromagnetic mode coupled to the quantum dot excitations (the next cavity mode is separated $\sim~$20 meV) \cite{Gerard}. Furthermore, we consider an homogeneous magnetic field along the z direction (the pillar growth direction) \cite{Whittaker}.

In our calculations, we take the lateral confinement in the quantum dot as  a parabolic potential. A Landau basis of single particle states for electrons (holes) is used. The parameters have been taken as follows. Effective in-plane masses $m_{e}=0.05~m_{0}$, $m_{h}=0.07~m_{0}$, where $m_{0}$ is the free electron mass, are used. $\hbar \omega_{0}=1 $meV  is the confinement energy. $\hbar\omega=1$ meV is the photon energy, measured with respect to the nominal band gap. The photon-matter coupling strength is given by  $g=0.5~$meV. We take the Coulomb interaction constant $\beta=2.73~\sqrt{B}~$meV, and the cyclotronic frequencies for electrons and holes as $\hbar \omega_{c}^{e(h)}=1.15*10^{-1}B/m_{e(h)}~$ meV, where  B is given in Teslas.
The Hamiltonian for this problem is 

\begin{eqnarray}
\hat{H} &=&\sum_{n}(E_{n}^{(e)}e_{n}^{\dagger}e_{n}+E_{\bar{n}}^{(h)}h_{{\bar{n}}}^{\dagger}h_{\bar{n}})+\sum_{n}(t_{kn}^{(e)}e_{k}^{\dagger}e_{n}+t_{\bar{k}\bar{n}}^{(h)}h_{\bar{k}}h_{\bar{n}})\nonumber \\ &+&\frac{\beta}{2} \sum_{r,s,u,v}\left\langle r,s\right|\frac{1}{r}\left|u,v\right\rangle e_{r}^{\dagger}e_{s}^{\dagger}e_{v}e_{u}+\frac{\beta}{2} \sum_{r,s,u,v}\left\langle r,s\right|\frac{1}{r}\left|u,v\right\rangle h_{\bar{r}}^{\dagger}h_{\bar{s}}^{\dagger}h_{\bar{v}}h_{\bar{u}}\nonumber \\ &-&\beta \sum_{r,s,u,v}\left\langle r,s\right|\frac{1}{r}\left|u,v\right\rangle e_{r}^{\dagger}h_{\bar{s}}^{\dagger}h_{\bar{v}}e_{u} +\hbar \omega a^{\dagger}a+g\sum_{n}(a^{\dagger}e_{n}h_{\bar{n}}+ah_{\bar{n}}^{\dagger}e_{n}^{\dagger})
\end{eqnarray}

\section{$T=0$}
We use a BCS-variational method with a trial function given by  
\begin{equation}
\left|u_{n},v_{n},\sigma\right\rangle=\left|\sigma\right\rangle\otimes \prod^{Nstates}_{n}(u_{n}+v_{n}e_{n}^{\dagger}h_{n}^{\dagger})\left|0\right\rangle,
\end{equation}
\noindent
where $\left|\sigma\right\rangle$ is a coherent state for the photons, and $\prod^{Nstates}_{n}(u_{n}+v_{n}e_{n}^{\dagger}h_{n}^{\dagger})\left|0\right\rangle$ describes the electron-hole pairs.
\noindent
The number of polaritons  is defined as $ \hat{N}_{pol}=\hat{N}_{pairs}+\hat{N}_{ph}$. 
By minimizing the $\langle \hat{H} \rangle$ with respect to $\sigma$ and $v_{n}$ we get the following equations

 \begin{eqnarray}
(\hbar \omega-2\mu_{ex})\sigma-g\sum_{i}\frac{1}{2}\frac{\Delta_{i}}{\sqrt{\Delta_{i}^2+(\epsilon_{i}-\mu_{ex})^2}}=0, \label{eq3}
\end{eqnarray}

\begin{equation}
\Delta_{k}=\beta \sum_{i,i\neq k}[i:k]\frac{1}{2}\frac{\Delta_{i}}{\sqrt{\Delta_{i}^2+(\epsilon_{i}-\mu_{ex})^2}}+g\sigma \label{eq4},
\end{equation}
\noindent
the latter is a generalized gap equation. $[i:j]=\langle ij \mid \frac{1}{r} \mid ji \rangle$. The explicit expression for the mean number of polaritons is
\begin{eqnarray}
N_{pol}=\vert\sigma \vert^2+\frac{1}{2}\sum_{i}\left(1-\frac{(\epsilon_{i}-\mu_{ex})}{\sqrt{\Delta_{i}^2+(\epsilon_{i}-\mu_{ex})^2}}\right),\label{eq5}
\end{eqnarray}

\begin{figure}[ht]
\begin{center}
\includegraphics[width=0.9\linewidth,angle=0]{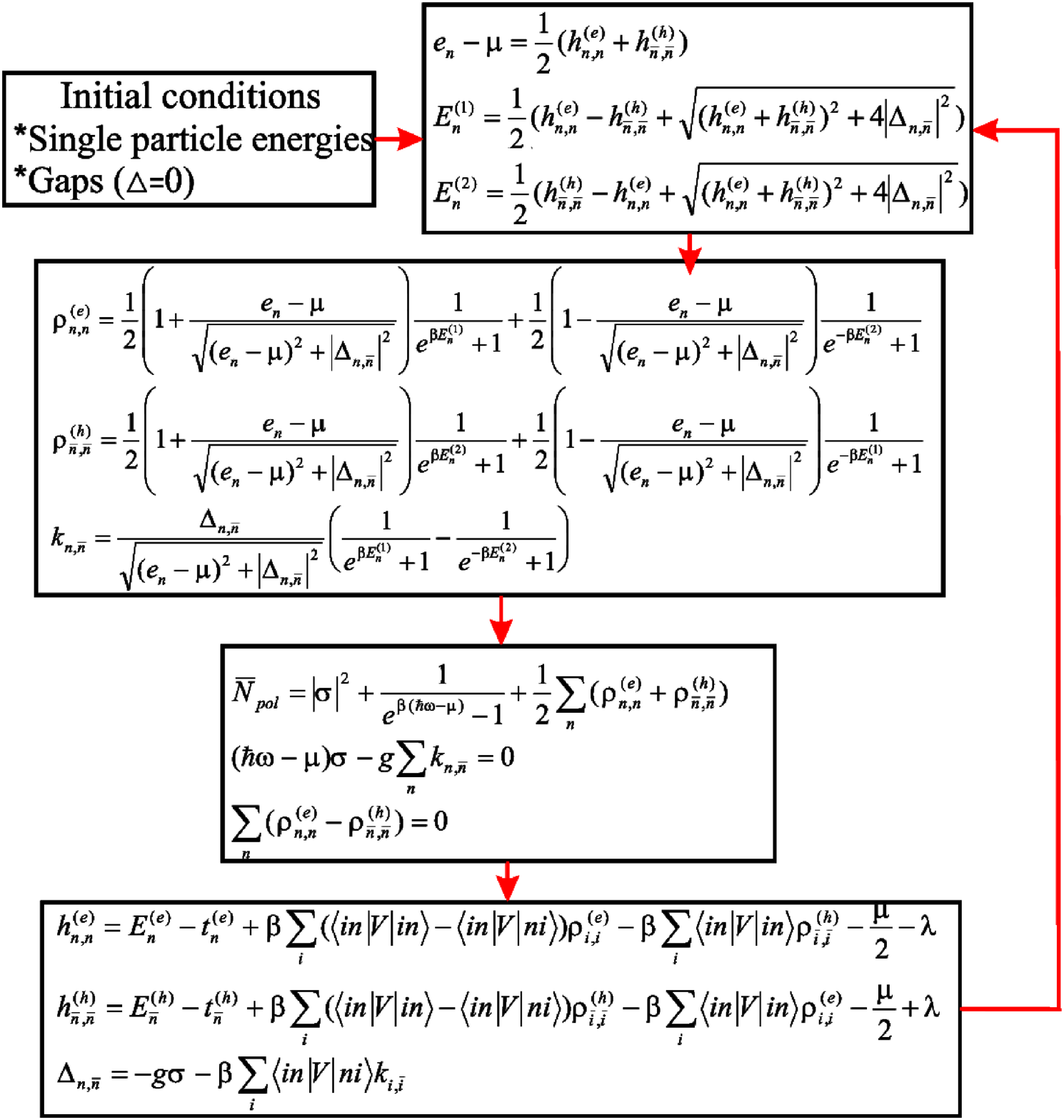}
\caption{\label{fig1} Selfconsistent procedure at finite temperatures.}
\end{center}
\end{figure}

where we used the standard parametrization for the occupations:

\begin{equation}
v_{k}^{2}=\frac{1}{2}\left(1-\frac{(\epsilon_{i}-\mu_{ex})}{\sqrt{\Delta_{i}^2+(\epsilon_{i}-\mu_{ex})^2}}\right),
\end{equation}
$\mu_{ex}$ (chemical potential) is the Lagrange multiplier associated  to the conservation  of $N_{pol}$, and $\epsilon_{i}$ is the pair energy.
\begin{eqnarray}
\epsilon_{i}&=&\frac{1}{2}(E_{i}^{(e)}+E_{\bar{i}}^{(h)}+t_{i,i}^{(e)}+t_{\bar{i},\bar{i}}^{(h)})-\frac{\beta}{2}[i:i]\nonumber\\&-&\beta \sum_{j,(j\neq i)}[i:j]v_{j}^{2},
\end{eqnarray}
Eqs. (\ref{eq3}-\ref{eq5}) are solved iteratively to obtain  $\sigma$, $\mu_{ex}$ and the gap parameters $\Delta_{i}$.

\section{$T\neq0$}
At finite temperatures, the starting point is the density matrix \cite{HK,Blaizot,Fetter}, 
\begin{equation}
D=\frac{e^{-\beta K_{ph}}}{Z_{ph}}\frac{e^{-\beta K_{eh}}}{Z_{eh}}
\end{equation}
 Where $Z_{ph}=Tr e^{-\beta K_{ph}}$, $Z_{eh}=Tr e^{-\beta K_{eh}}$, $K_{ph}=\hbar \omega-\mu_{ex} $, and $K_{eh}$ is expressed in terms of single- particle energies and gap functions. The thermodynamical potential 
\begin{equation}
\phi(D)=Tr(DH)-\mu \bar{N}_{pol}+\frac{1}{\beta}Tr(D~lnD)
\end{equation}
Should be minimized with respect to $\sigma$ and the $v_{n}$. The selfconsistent
 procedure is schematically sketched in  Fig.\ref{fig1}.

\section{NUMERICAL RESULTS}
\begin{figure}[h]
\begin{center}
\includegraphics[width=.8\linewidth,angle=0]{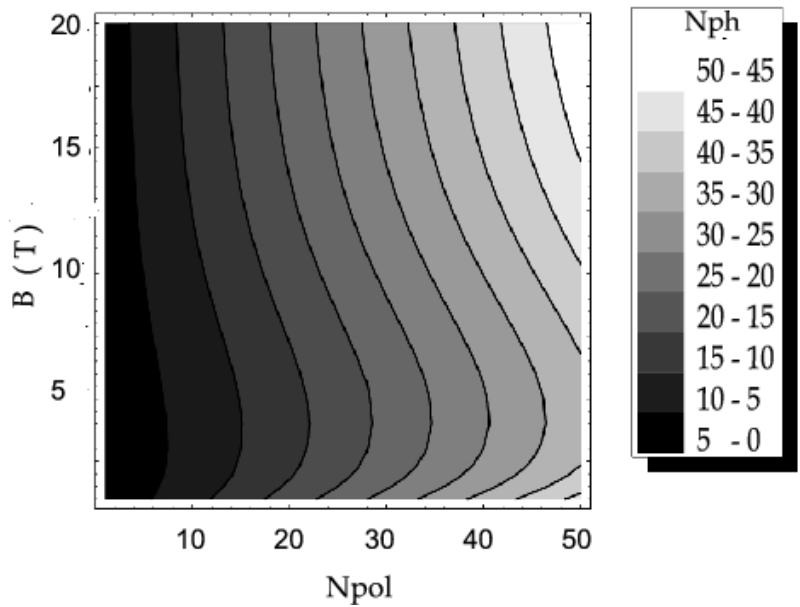}
\caption{\label{fig2} The number of photons as a function of the number of polaritons and the magnetic field, $\hbar \omega=1$ meV and T=0.}
\end{center}
\end{figure}
At $T=0$, we show in Fig. \ref{fig2} the mean number of photons in the cavity as a function of B and the number of polaritons. At fixed polariton number we obtain a non monotonous dependence on B, which can be understood in terms of the effect of B on the pair energies. At low B, the energies decrease because of a term of the form of $\sim\frac{\hbar e}{2}(\frac{1}{m_{e}}+\frac{1}{m_{h}})B\langle l_{e}\rangle$, where the mean value of the electron angular momentum becomes negative. Thus the detuning ($E_{photon}$-$E_{pair}$) increases and the mean number of photons decreases. On the other hand, for large values of B, the contribution of the lowest Landau  level to the energy, $\frac{\hbar e}{2}(\frac{1}{m_{e}}+\frac{1}{m_{h}})B$, makes the detuning lower, thus augmenting the number of photons. 
\begin{figure}[ht]
\begin{center}
\includegraphics[width=.7\linewidth,angle=0]{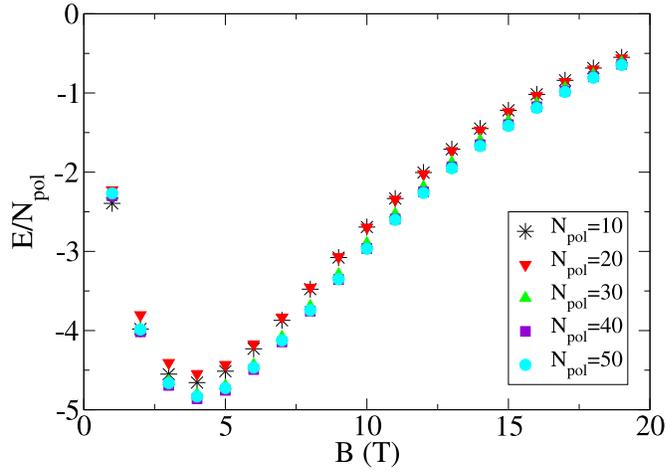}
\caption{\label{fig3} Scaling of the total energy, $T=0$ and $\hbar \omega=1$ meV.}
\end{center}
\end{figure}
\begin{figure}[ht]
\begin{center}
\includegraphics[width=0.6\linewidth,angle=0]{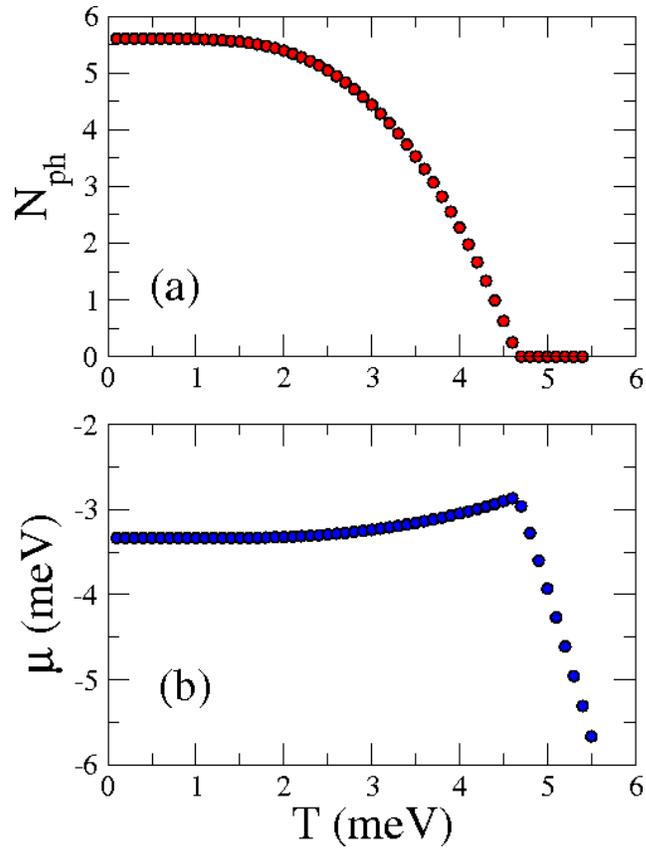}
\caption{\label{fig4} The number of photons as a function of temperature for
  $\hbar \omega=1 meV$, $\hbar \omega_{0}=1$ meV and $N_{pol}=10$. }
\end{center}
\end{figure}

The total energy as a function of B is drawn in Fig. \ref{fig3}. The change in the slope from low B to large B has the same origin as discussed above. Coulomb interactions at large B values gives a contribution $\sim-\sqrt{B}$ which in conjunction with the dominant term from the lowest Landau level makes the curve concave. The scaling $E\sim N_{pol}$ could be taken as an indication that the effective interactions between quasi-particles (polaritons) is very weak (we expect a contribution $\sim N_{pol}^{2}$ coming from interactions). The scaling is completely natural for high B, as it happens for excitons \cite{Lozovik}.

In  Fig. \ref{fig4}. We show the calculated mean number of photons in the condensate $(\mid \sigma \mid ^{2})$ and the chemical potential as a function of temperature for a 10-polariton system at $B=7~$Teslas. The transition at $T\approx4.6~$meV is interpreted  as Bose condensation of polaritons in the same way that it happens for the Dicke model \cite{Littlewood}. Notice that, as follows from our equations  (see Fig. \ref{fig1}), a zero value for $\sigma $ implies zero gap functions $\Delta_{n,\bar{n}}$.

\begin{figure}[h]
\begin{center}
\includegraphics[width=0.6\linewidth,angle=0]{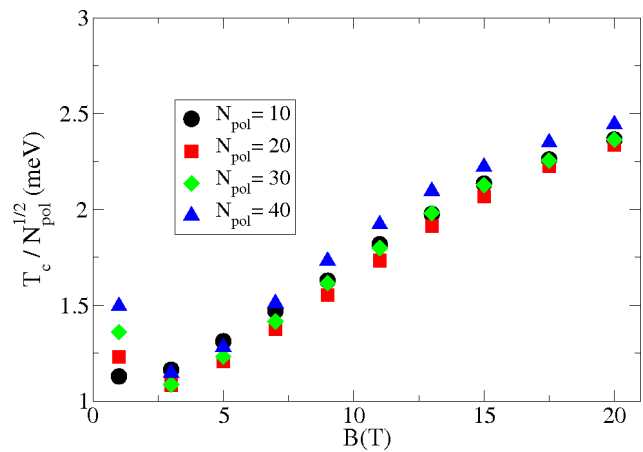}
\caption{\label{fig5} Critical temperature vs magnetic field.}
\end{center}
\end{figure}

The results of more extensive calculations for different $N_{pol}$ numbers and magnetic fields are represented in Fig.\ref{fig5}. The dependence $T_{c}\sim N_{pol}^{1/2}$ is typical of free 2D bosons in an harmonic potential. Thus, the approximate scaling with $N_{pol}^{1/2}$ again reinforces  the picture of almost noninteracting effective quasi-bosons (polaritons). The obtained values of $T_{c}$ are of the order of the pair binding energies,
$E_{n}^{(e)}+t_{n,n}^{(e)}+E_{\bar{n}}^{(h)}+t_{\bar{n},\bar{n}}^{(h)}-\epsilon_{n}$, and behave  like $\sqrt{B}$ at large B. An apparent change of slope at $B\approx3~$Teslas could be an artifact of our finite basis.

In conclusion, we presented a theoretical study of a quantum dot-microcavity system  in an homogeneus magnetic field at zero and finite temperatures.
 We showed that the magnetic field can be used as a control parameter to vary the mean number of photons in the cavity and the critical temperature for BE condensation. The scaling of the total energy and critical temperature with $N_{pol}$ indicates that the effective quasi-particles in the system are weakly interacting quasi-bosons.

The authors acknowledge support by the Comitee for Research of the Universidad
de Antioquia and the Caribbean Network for Quantum Mechanics, Particles and Fields (ICTP).


\begin{thebibliography}{99}
\bibitem{X1} J.M. Gerard, B. Sermage, B. Gayral, B. Legrand, E. Costard,
 and V. Thierry-Mieg, Phys. Rev. Lett. {\bf 81}, 1110 (1998).
\bibitem{X2} G.S. Solomon, M. Pelton, and Y. Yamamoto, Phys. Rev. Lett.
 {\bf 86}, 3903 (2001).
\bibitem{X3} M. Pelton, C. Santori, J. Vuckovic, B. Zhang, G.S. Solomon,
 J. Plant, and Y. Yamamoto, Phys. Rev. Lett. {\bf 89}, 233602 (2002).
\bibitem{Dang} J.Kasprzak, M. Richard, S. Kundermann,A. Baas,P. Jeambrun, J.M.J. Keeling, F.M. Marchetti, M.H. Szymanska, R. Andr\'e, J.L. Staehli, 
V. Savona, P.B. Littlewood, B. Deveaud and Le Si Dang, Nature {\bf 443}, 05131 (2006).
\bibitem{Yamamoto2006} H. Deng, D. Press, S. G\"otzinger, G.S. Solomon, R. Hey, K.H. Ploog and Yoshihisa Yamamoto. Phys. Rev. Lett. {\bf 97} 146402(2006).
\bibitem{XX1} O. Benson, C. Santori, M. Pelton, and Y. Yamamoto, Phys. Rev. Lett.
 {\bf 84}, 2513 (2000).
\bibitem{Snocke} L.V. Keldysh, in Bose - Einstein Condensation, edited by A. Griffin, D.W. Snocke, and A. Stringari (Cambridge University Press, England, 1995).
\bibitem{Gerard}J.M. Gerard, D. Barrier,J.Y. Marzin, R. Kuszelewicz, L. Manin, E. Costard, V. Thierry-Mieg, T. Rivera, Appl. Phys. Lett. {\bf 69} (1996) 449.
\bibitem{Whittaker}A. Armitage, T.A. Fisher, M.S. Skolnick, D.M. Whittaker, P. Kinsler and J.S. Roberts. Phys . Rev. {\bf B, 55}, 16395 (1997).
\bibitem{HK} H. Haug and S.W. Koch, {\it Quantum Theory of the Optical
 and Electronic Properties of Semiconductors} (World Scientific,
 Singapore, 1994).
\bibitem{Blaizot} J.P. Blaizot and G. Ripka, Quantum Theory of Finite Systems. MIT Press, Cambridge, Mass., 1986.
\bibitem{Fetter} A. Fetter, J.M. Walecka, Quantum Theory of Many-Particle Systems, McGraw-Hill, New York, 1971.
\bibitem{Lozovik}I.V. Lerner and Yu. E. Lozovik, Zh. \'Eksp. Teor. Fiz. {\bf 80}, 1488 (1981) [Sov. Phys. JETP {\bf 53}, 763(1981)].
\bibitem{Littlewood} P.R. Eastham and P.B. Littlewood. Phys. Rev {\bf B, 64}, 235101 (2001).
\end{thebibliography}
\end{document}